\documentclass[aip,graphicx]{revtex4-2}
\usepackage{amsmath}
\usepackage{graphicx}
\usepackage{color}

\begin{document}

\title{Computing the multimodal stochastic dynamics of a nanobeam in a viscous fluid}

\author{J. Barbish}
\affiliation{Department of Mechanical Engineering, Virginia Tech, Blacksburg, Virginia 24061, USA}

\author{M. R. Paul}
\email{mrp@vt.edu}
\affiliation{Department of Mechanical Engineering, Virginia Tech, Blacksburg, Virginia 24061, USA}

\date{\today}

\begin{abstract}
The stochastic dynamics of small elastic objects in fluid are central to many important and emerging technologies. It is now possible to measure and use the higher modes of motion of elastic structures when driven by Brownian motion alone. Although theoretical descriptions exist for idealized conditions, computing the stochastic multimodal dynamics for the complex conditions of experiment is very challenging.  We show that this is possible using deterministic finite element calculations with the fluctuation dissipation theorem by exploring the multimodal stochastic dynamics of a doubly-clamped nanobeam. We use a very general, and flexible, finite-element computational approach to quantify the stochastic dynamics of multiple modes simultaneously using only a single deterministic simulation.  We include the experimentally relevant features of an intrinsic tension in the beam and the influence of a nearby rigid boundary on the dynamics through viscous fluid interactions. We quantify the stochastic dynamics of the first eleven flexural modes of the beam when immersed in air or water. We compare the numerical results with theory, where possible, and find excellent agreement.  We quantify the limitations of the computational approach and describe its range of applicability.  These results pave the way for computational studies of the stochastic dynamics of complex 3D elastic structures in a viscous fluid where theoretical descriptions are not available. 

\end{abstract}

\maketitle

\section{Introduction}
\label{sec:introduction}

The Brownian driven fluctuations of small elastic structures in fluid represent the ultimate limit in detecting small scale motions at a finite temperature for many systems of intense interest. Approaches exploiting these fluctuations have yielded exciting new technologies with unprecedented capabilities~\cite{arlett:2011,ekinci_nanoelectromechanical_2005,xu:2022:nano,bachtold:2022}. Examples include probing the dynamics of a single molecule tethered between cantilevers with piconewton force resolution and microsecond temporal sensitivity~\cite{paul_stochastic_2004,radiom:2013}, detecting the signature of life~\cite{kasas:2015,roslon:2022} or the efficacy of antibiotics~\cite{longo:2013,lissandrello:2014,villalba:2021} by measuring the change in the fluctuations of an elastic object when a living sample or target bacterium is attached, quantifying the microrheology of complex fluids~\cite{ma:2000,radiom:2012,nishi:2021}, measuring the fluctuations of a carbon nanotube in a liquid~\cite{sawano:2010,barnard:2019}, and the development of stochastic heat engines that use mechanical vibrations to harvest energy between thermal baths~\cite{serra-garcia:2016}, to name just a few.

Many approaches use micro and nanoscale cantilevers and doubly-clamped beams as the Brownian driven elastic object~\cite{ekinci_nanoelectromechanical_2005,clarke:2006,honig:2012}. It is now possible to accurately measure the multimode stochastic dynamics of these structures in fluid~\cite{gress_multimode_2023}.  An advantage of the high mode-number dynamics is the increased frequencies of the resonant peaks, yielding improved temporal resolution. In addition, the frequency variation of the added mass and viscous damping of the fluid results in an increase of the quality factor with increasing mode number which can be very beneficial by yielding well defined peaks in the amplitude frequency response. It is anticipated that measurements of the stochastic dynamics of the higher modes of vibration will continue to be exploited in future technologies.

Significant progress has been made exploring the multimodal dynamics of small beams and elastic structures when driven externally with a harmonic drive including electrothermal actuation~\cite{bargatin:2007,ma:2023}, piezoelectric actuation~\cite{ghatkesar:2008}, and pump-probe approaches~\cite{pelton:2013,yu:2015}.  Computational efforts investigating harmonically driven objects have explored the hydrodynamics of the fluid-solid interactions~\cite{aureli:2012,phan:2013,ricci:2013,gesing:2022,shen:2023,devsoth:2024}. However, we emphasize that the dynamics of an externally driven elastic object and the dynamics of a Brownian driven object are significantly different in important and interesting ways~\cite{paul_stochastic_2006}. Typically, for a system being driven externally, the magnitude of the driving force is chosen to be sufficiently large that the measured displacements are much larger than the thermally driven displacements, resulting in an increase in the signal to noise ratio. Even when considered in the frequency domain, the spectral properties of the elastic objects significantly depend upon the nature of the drive~\cite{sader_frequency_1998,paul_stochastic_2006,clark_spectral_2010}.  For example, the frequency variation of the amplitude spectra and the diagnostic quantities, such as the quality factor and the resonant frequency of the different modes, depend upon the nature of the driving force. The computational exploration of the multimodal stochastic dynamics requires a different approach than what is used for systems driven externally with a harmonic drive.

In light of the favorable characteristics of the multimodal stochastic dynamics of elastic structures, we explore a computational approach capable of quantifying the stochastic dynamics for the complex geometries and conditions that describe many experimental conditions where a theoretical description is not currently available. A powerful approach for computing the stochastic dynamics of elastic objects in fluid~\cite{paul_stochastic_2004,paul_stochastic_2006} relies upon the fluctuation-dissipation theorem~\cite{callen_irreversibility_1951,callen_theorem_1952}. In this numerical approach, a single deterministic calculation can be used to yield the stochastic dynamics. In essence, the important calculation is to compute the system's \emph{deterministic} return to equilibrium after a prescribed excursion from equilibrium.

For fluctuations in the displacement of an elastic object in a fluid, this is the removal of a force, applied in the distant past, and computing the ring-down of the object as it returns to a quiescent equilibrium state. This approach is very flexible and is valid for all of the modes of the elastic structure. If the experimental system can be described numerically, such as the fluid-solid interaction of a beam in a viscous fluid, then the powerful deterministic numerical approaches that are available, such as finite element methods, can be used.  This approach has been very successful at describing the stochastic dynamics of the fundamental mode of elastic structures over a wide range of conditions, for example see Refs.~\cite{paul_stochastic_2006,clark_spectral_2010,honig:2012}.

There are several important advantages to this numerical approach which we exploit in our current study.  An important point to highlight is that the finite-element computations include the contributions from all of the modes in the elastic structure. Specifically, no expansion in modes is necessary which is typical of analytical descriptions. As a result, the numerical approach rigorously describes the dynamics in the case of overlapping modes for low quality oscillators which is common when fluids such as water are used. In addition, the numerical approach includes any couplings that may exist between the modes due to the subtle fluid-solid interactions. A very powerful feature of the computational approach is that it includes the full three dimensional fluid dynamics. This includes the additional axial flows that are expected to occur at high mode number~\cite{van_eysden_frequency_2007} in addition to the influence of the bounding walls of the fluid domain on the beam dynamics through viscous interactions~\cite{clarke:2006,tung_hydrodynamic_2008,clark:2008}. Therefore, the computational approach provides useful insights into the accuracy of available theories and their ability to predict the multimodal dynamics in fluid. The computational approach can be used to quantify the influence of contributions such as axial fluid flows and mode couplings for a broad range of conditions.

In this paper we demonstrate that computations, using this fluctuation-dissipation theorem based approach, can be extended to describe the multimodal stochastic dynamics of elastic structures for experimentally realistic conditions.  We compute the stochastic dynamics of the first eleven modes of a doubly-clamped beam immersed in air or water.  We include an intrinsic tension in the beam and the influence of a nearby rigid wall on the stochastic dynamics through interactions of the viscous fluid.  These features were included with the typical conditions of experiments~\cite{gress_multimode_2023} in mind. However, we emphasize the generality of the computational approach and the straightforward nature of including other physical features as well. For example, additional features that could be included in the computations are a slip boundary condition, the added mass of an attached biomolecule, or the inclusion of additional sources of damping that may be present in the system.  A hallmark of the computational approach is the ability to quantify the entire frequency response, including the multimodal dynamics, from a single deterministic calculation.

The remainder of the paper is organized as follows. We first discuss the physical problem that we explore in Sec.~\ref{sec:prelimiary-discussion}.  We next describe the computational approach and its relationship to the fluctuation-dissipation theorem in Sec.~\ref{sec:computational-approach}. In Sec.~\ref{sec:results} we discuss our investigation of the stochastic multimodal dynamics of a doubly clamped beam that is immersed in water or air. Lastly, in Sec.~\ref{sec:conclusions} we provide our concluding remarks.

\section{Preliminary Discussion}
\label{sec:prelimiary-discussion}

\begin{figure}[h!]
\begin{center}
\includegraphics[width=3.25in]{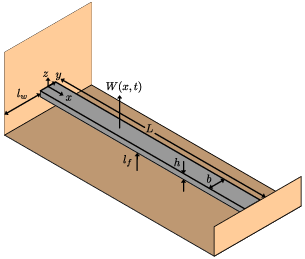}
\end{center}
\caption{A doubly-clamped beam with length $L$, width $b$, and height $h$. The geometric and material properties of the gray beam are listed in Table~\ref{tab:beam}. The beam has a nearby floor, a distance $l_f$ beneath the bottom surface of the beam. The remaining three walls (not shown) are a distance of $l_w$ away from the corresponding beam surface.}
\label{fig:beam-diagram}
\end{figure}

We consider a doubly-clamped beam of length $L$, width $b$, and thickness $h$ that is immersed in a viscous fluid as shown in Fig.~\ref{fig:beam-diagram}. The beam, shown in gray, is long and thin such that $L \! \gg \! b \! \gg \! h$. The beam is clamped to solid walls at its two distal ends. The beam fluid interfaces are no-slip surfaces. The floor of the fluid domain is a distance $l_f$ from the bottom surface of the beam and the remaining three bounding walls (not all shown for clarity) are a distance $l_w$ from the nearest surface of the beam. All six bounding surfaces of the fluid domain are no-slip surfaces. The axial coordinate along the beam is $x$, the transverse coordinate is $z$, and the remaining in-plane coordinate is $y$. We will follow the convention that the axial coordinate $x$ has been nondimensionalized by $L$  such that $x\!=\!0$ is the left side of the beam and $x\!=\!1$ is the right side of the beam. The transverse displacement of the beam at position $x$ and time $t$ will be represented as $W(x,t)$. 

The specific geometry and material parameters of the nanobeam we explore are given in Table~\ref{tab:beam}. The nondimensional tension parameter,  $U\!=\! \frac{F_T L^2}{2 E I}$, relates the tension force to an elastic force scale. For $U \gtrsim 12$ the beam can be considered to be under high tension~\cite{bokaian_natural_1990}.  The values of Table~\ref{tab:beam} are chosen to be similar to currently used nanobeams in experiments~\cite{gress_multimode_2023}. However, it is not our intention here to explore the specifics of any particular experiment but rather to present a broader study demonstrating the generality of the approach while providing new physical insights into the multimodal stochastic dynamics for experimentally relevant conditions. We quantify the stochastic dynamics of the beam when immersed in air or water. The fluid properties and details of the fluid domain are given in Table~\ref{tab:fluid-prop}.
\setlength{\tabcolsep}{12pt}
\begin{table}[h!]
\begin{center}
\begin{tabular}{ c c c c c c c} 
 $L$ & $b$ & $h$ & $\rho_s$ & $E$ & $F_T$ & $U$ \\
 $\left[\mu\text{m}\right]$ & [$\mu$m] & [nm] & [kg/m$^3$] & [GPa] & [$\mu$N] & \\ \hline \hline 
 40 & 0.90 & 93 & 2960 & 300 & 8.5 & 375.7 \\ 
\end{tabular}
\end{center}
\caption{Geometry and material properties of the nanobeam. The beam has length $L$, width $b$, and thickness $h$. The density $\rho_s$, Young's modulus $E$, and tension force $F_T$ are chosen to be close to experimentally relevant values for a silicon nitride beam. U is the nondimensional tension parameter.}
\label{tab:beam}
\end{table}

In many cases, it is desired to choose the size of the fluid domain to be large enough such that the bounding walls do not influence the dynamics of the elastic object through viscous interactions.  It is often desirable to not have the size of the fluid domain exceed this value in order to avoid unnecessary computational expense. The appropriate size of the fluid domain to use in the computations depends upon the fluid used and upon the dynamics of the nanobeam. The size of the fluid domain can be estimated using the Stokes length, $\delta_s(\omega)$, as a guide where $\delta_s$ is the effective thickness of the unsteady viscous boundary layer and $\omega$ is the frequency of oscillation~\cite{rosenhead_laminar_1963}. This can be quantified approximately using the Stokes length $\delta_s(\omega) = (\nu_f/\omega)^{1/2}$ where $\nu_f = \mu_f/\rho_f$ is the kinematic viscosity of the fluid. Using the frequency of the fundamental mode in fluid, $f_{p,1}^{(t)} = \frac{\omega_{p,1}^{(t)}}{2 \pi}$, yields the values of $\delta_{s,1}$ shown in Table~\ref{tab:fluid-prop}. In our notation, a subscript of $p,n$ on a frequency indicates the frequency of the peak of the amplitude spectrum for mode $n$ of a beam immersed in fluid. A subscript of $s,n$ on $\delta$ indicates the Stokes length computed using the frequency of mode $n$, and a superscript $(t)$ indicates a theoretical value.

Finally, the distance from the beam to the walls of the domain, $l_w$, is chosen such that the walls are far away and do not significantly affect the dynamics. In our study we have used $l_w/\delta_{s,1} \approx 10$. We note that $\delta_{s,1} \!>\! \delta_{s,2} \!>\! \!\ldots\! \!>\! \delta_{s,n}$ and therefore it is only necessary to consider the fundamental mode when determining the size of the fluid domain.  The distance $l_f$ from the bottom beam surface to the wall below, which we will refer to as the floor, as shown in Fig.~\ref{fig:beam-diagram}, will be varied as part of our study. In experiment, the floor is often the closest boundary to the oscillating beam and its presence can affect the dynamics.
\begin{table}[h!]
\begin{center}
\begin{tabular}{ c c c c c c}
fluid & $\rho_f$ & $\mu_f$ & $f_{p, 1}^{(t)}$ & $\delta_{s, 1}$ & $l_w$\\
 & $\left[\frac{\text{kg}}{\text{m}^3}\right]$ & $\left[\frac{\text{kg}}{\text{m} \, \text{s}} \right]$  & [MHz] & [$\mu$m]  & [$\mu$m]\\ \hline \hline 
air & 1.23 & $1.79\times 10^{-5}$  & 2.486 & 0.97 & 10\\
water & 997.8 & $9.53\times 10^{-4}$  & 0.556 & 0.52 & 5\\
\end{tabular}
\end{center}
\caption{Fluid properties of air and water. The fluid density $\rho_f$, dynamic viscosity $\mu_f$, the theoretical prediction of the frequency of the first peak in noise spectrum $f_{p, 1}^{(t)}$, and the Stokes length $\delta_{s,1}$ where  $l_w/\delta_{s, 1} \sim 10$ for both fluids.}
\label{tab:fluid-prop}
\end{table}

Although it is well understood that the dynamics of the fundamental mode of oscillation of typical micro and nanoscale beams can be described using continuum ideas, such as the Euler-Bernoulli beam equation and the Navier-Stokes equations of fluid dynamics, these approximations will eventually become invalid as the length and time scales become smaller. This is of particular interest when considering the higher modes of micro and nanoscale structures which occur at higher frequencies and with smaller displacements.  In our study, we explore the first eleven modes of the nanobeam and, in this light, we briefly discuss the length and time scales under consideration in more detail to establish the range and limitations of the continuum assumption.

The Knudsen number, $\text{Kn} \!=\! \lambda_f/L_b$, is the ratio of the molecular length scale to the length scale of the elastic object where $\lambda_f$ is the mean free path between collisions of the molecules in the fluid and $L_b$ is a characteristic length scale of the beam. $\text{Kn} \ll1$ is an indicator that the continuum hypothesis is appropriate from a length scale perspective. For an oscillating beam in fluid, $L_b$ is typically chosen as the beam width $b$ since the bulk of the fluid motion is around the sides of the beam as it oscillates in the transverse direction. In fact, it is for this reason that a 2D description of the fluid dynamics often suffices for a theoretical description for oscillating beams~\cite{sader_frequency_1998}.

However, for the higher modes of the beam it is important to also consider the axial dependence of the beam deflections which can lead to three-dimensional fluid flows. This can be quantified using the mode dependent spatial wavelength, $L_n$, as $L_b$ to yield a mode dependent Knudsen number $\text{Kn}_n \!=\! \lambda_f/L_n$.  For a doubly-clamped beam, $L_n$ can be estimated as $L_n \!=\! L/n$ where $n$ is the mode number such that $\text{Kn}_n \!=\! n \lambda_f / L$. For the slender beam we consider, $L/b \!=\! 44.4 \!\gg\! 1$, this yields $\text{Kn}_{11} \!<\! \text{Kn}$ and therefore $\text{Kn}$ is a good indicator of the ratio of the important length scales for the dynamics we explore.

When the beam described in Table~\ref{tab:beam} is immersed in water, this yields $\text{Kn} \!=\! 3.3 \! \times \! 10^{-4}$ where we have used the diameter of a water molecule as an estimate of the mean free path such that $\lambda_f \!\approx\! 0.3$ nm.  This indicates that the continuum hypothesis is valid for all of the beam dynamics we explore in water. When the beam is immersed in atmospheric air, $\text{Kn} \!=\! 0.11$, where $\lambda_f \!\approx\! 100$ nm. Therefore, even in atmospheric air the Knudsen number remains small for all of the dynamics we explore. Using these scaling arguments, the mode number $n_c$ for which $\text{Kn}_{n_c} \! \approx \! 1$, where molecular effects would be expected to dominate is $n_c \! \approx \! 400$ which is much larger than what has been explored.

The continuum hypothesis also requires that the time scale of the dynamics under investigation is much larger than the time scale of molecular relaxation in the fluid. This is captured by the Weissenberg number, $\text{Wi} \!=\! \tau_f/\tau_{b}$, where $\tau_f$ is the relaxation time of the fluid and $\tau_b$ is a characteristic time scale of the beam dynamics. $\text{Wi} \!\ll\! 1$ is an indicator that the continuum hypothesis is justified from a time scale point of view. Since we are interested in the dynamics of the higher modes, we use a generalized and mode-dependent Weissenberg number, $\text{Wi}_n \!=\! \tau_f/\tau_{b,n}$, where $\tau_{b,n}$ is the characteristic time scale of mode $n$. We use the period of oscillation of mode $n$ as the time scale, $\tau_{b,n} = 1/f_{p,n}$, where $f_{p,n}$ is the frequency of oscillation of mode $n$, when immersed in fluid, which yields $\text{Wi}_n \!=\! \tau_f f_{p,n}$. In our notation, $f_n$ is the natural frequency and $f_{p,n}$ is the resonant frequency of the mode when the beam is immersed in fluid where $f_{p,n} < f_n$.

The thermal relaxation time of room temperature water is extremely short which can be estimated as $\tau_f \approx 1$ ps. When the beam is immersed in water, the peak frequency of oscillation for the different modes reduces significantly from their natural values due to the mass loading and viscous damping of the fluid. In this case, we use the peak frequency of the noise spectrum for $f_{p,n}$ used in the $\text{Wi}_n$.  For the beam we explore, $f_{p,1} \!\approx\! 550$ KHz which yields $\text{Wi}_1 \!=\! 5.5 \!\times\! 10^{-7}$ and $f_{p,11} \!\approx\! 19$ MHz which yields $\text{Wi}_{11} \!=\! 1.9 \times 10^{-5}$. Therefore, for the beam in water, we have $\text{Wi}_n \!\ll\! 1$ for all of the conditions we explore.

When the beam is immersed in air, the frequencies do not change significantly from their natural values and an estimate of the relaxation time is $\tau_f \approx 0.8$ ns~\cite{kara_generalized_2017,jennings_mean_1988}. In this case, $f_1 \!=\! 2.5$ MHz and $f_{11} \!=\! 43$ MHz, which yields $\text{Wi}_{1} \!=\! 2 \!\times\! 10^{-3}$ and $\text{Wi}_{11} \!=\! 0.034$. Therefore, for the conditions we explore the continuum approximation is valid in terms of the time scales. In air, the period of oscillation of a higher mode of the beam, $f_n^{-1}$, would not equal the relaxation time of the fluid $\tau_f$ until $f_n \!\approx\! 1/\tau_f \!=\! 1.25$ GHz which would not occur until mode number $n \!\approx\! 68$.

Kara~\emph{et al.}~\cite{kara_generalized_2017} performed a careful study over a wide range of conditions to characterize the transition from the molecular to continuum limit for oscillating objects of finite size and found that the continuum approximation is valid when $\text{Wi} \!+\! \text{Kn} \lesssim 1$. It is important to emphasize that the functional form of this condition as the sum of the Weissenberg and Knudsen numbers, as opposed to a more complicated expression in terms of $\text{Wi}$ and $\text{Kn}$, has been demonstrated experimentally and justified from a very general theoretical analysis~\cite{kara_generalized_2017}. For our slender beam, we generalize this expression to include the higher mode dynamics as $\text{Wi}_{11} \!+\! \text{Kn} \!\lesssim\! 1$. When the beam is immersed in water this yields $\text{Wi}_{11} \!+\! \text{Kn} \!=\! 3.5 \!\times\! 10^{-4}$ and when the beam is immersed in air this yields $\text{Wi}_{11} \!+\! \text{Kn} \!=\! 0.15$.  This also suggests that the continuum hypothesis is valid for our study when the fluid is water. For air, this suggests that the continuum hypothesis is valid however the beam dynamics is much closer to approaching the transition to the molecular regime.

\section{Computational Approach}
\label{sec:computational-approach}

We are interested in the multimodal stochastic dynamics of the fluctuations in beam displacement of a doubly-clamped beam that is under tension, immersed in a viscous fluid, and near a solid boundary. This computational approach has been very successful at describing the dynamics of the fundamental mode of elastic structures over a wide range of conditions including rectangular~\cite{paul_stochastic_2006} and V-shaped~\cite{clark:2008} cantilevers, pairs of cantilevers~\cite{clark:2007,paul_stochastic_2004, honig:2012,robbins:2014}, a complex shaped nanobeam near a microdisk optical resonator~\cite{epstein:2013}, and doubly-clamped beams~\cite{villa_stochastic_2009}. However, in all of these studies the focus of the investigation was on the stochastic dynamics of the fundamental mode.  In this paper, we generalize the computational approach to explore the multimodal stochastic dynamics of a doubly clamped-beam.

The deterministic component of this calculation is often referred to as the system's ring-down. For an extended structure like a beam, the removal of the force $F_0$, and the subsequent measurement of the ring-down can occur at any axial location $x$ along the the beam. However, the resulting analysis is straightforward if the force is applied at the same location $x_0$ where the beam displacement is measured. This is the approach we use here and we will discuss numerical results using two different axial locations. 

When a force $F_0$, applied to the beam in the distant past at location $x_0$, is removed at time $t\!=\!0$,  the deterministic variation of the beam displacement at $x_0$ as it returns to equilibrium is $W(x_0, t)$.   The autocorrelation of equilibrium fluctuations in the stochastic beam displacement $w(x_0,t)$ can be computed from this deterministic ring-down, $W(x_0,t)$, from~\cite{paul_stochastic_2004}  
\begin{equation}
    \langle w(x_0, 0) w(x_0, t) \rangle = \frac{k_B T}{F_0} W(x_0, t)
    \label{eq:auto}
\end{equation}
where $k_B$ is Boltzmann's constant, $T$ is the temperature, and the angle brackets indicate an equilibrium ensemble average.  In all of our results we will assume room temperature $T\!=\!298$ K.  In our notation, lower case $w(x,t)$ indicates the stochastic fluctuations in beam displacement and upper case $W(x,t)$ indicates the deterministic ring-down of the beam displacement after the removal of the force.  The power spectral density, or noise spectrum, of the beam displacements at axial position $x_0$ is given by~\cite{paul_stochastic_2004} 
\begin{equation}
    G_W(x_0, \omega) = 4 \int_{0}^{\infty} \langle w(x_0, 0) w(x_0, t) \rangle \cos(\omega t) dt
\label{eq:noise}
\end{equation}
where the subscript $W$ indicates that this noise spectrum is in terms of the beam displacement. In our approach, we compute $W(x_0,t)$ deterministically and then use Eqs.~(\ref{eq:auto})-(\ref{eq:noise}) to compute the stochastic dynamics of the fluctuating displacement in the time and frequency domains, respectively. It is important to highlight that the fluctuating displacement of the beam displacement, $w(x_0,t)$, is an extremely complicated function which is never computed directly. However, using the fluctuation-dissipation theorem allows for the \emph{exact} computation of the important statistical quantities of the autocorrelation and the power spectral density from the deterministic ring-down $W(x_0, t)$. 

We compute the deterministic ring-down of the beam using a finite element approach. The fully coupled fluid-solid interaction problem was solved using standard algorithms~\cite{chung_time_1993,jansen_generalized_2000} available in the commercial software Comsol~\cite{comsol}. If sufficient spatial and temporal resolution are used, the deterministic ring-down $W(x_0,t)$ contains the contributions from multiple modes of the beam and the fluctuation-dissipation theorem based approach remains valid for computing the stochastic dynamics of the higher modes.  Therefore, it is important to carefully determine the necessary spatial and temporal resolution of the deterministic calculation to capture the multimodal dynamics of interest.

The following steps were used to ensure sufficient spatial and temporal resolution for our finite element simulations to capture the first eleven modes of the beam~\cite{barbish:2023:thesis}. We describe our approach in general terms such that the overall procedure we used could be readily applied to other problems of interest. We first computed the natural frequencies and static displacements of the beam which were directly compared with theory~\cite{bokaian_natural_1990,barbish:2022}. For the solid mechanics of the beam, we found that using a maximum length scale of the beam's mesh to be on the order of the beam's thickness, $h$, provided sufficient resolution. This resulted in $\sim 2 \!\times\! 10^4$ beam elements with a typical finite element volume of $\sim 0.2 h^3$. This yielded an error of less than 0.3\% for the natural frequencies of the first eleven modes when compared with theory. The computed natural frequencies, $f_n^{(s)}$, and the theoretical values, $f_n^{(t)}$, are given in Table~\ref{tab:fn-kn}. In our notation, a  superscript $(s)$ indicates a value from a numerical simulation. The theoretical values of the natural frequencies are determined using the theory described in detail in Ref.~\cite{barbish:2022}.
\setlength{\tabcolsep}{12pt}
\begin{table}[h!]
\begin{center}
\begin{tabular}{ c c c c c } 
 $n$ & $f_n^{(t)}$ & $f_n^{(s)}$ & $k_n(x_0\!=\!1/2)$ & $k_n(x_0\!=\!1/4)$ \\
  & [MHz] & [MHz] & [N/m] & [N/m] \\ \hline \hline 
1 & 2.514 & 2.517 & 1.150 & 2.616 \\ 
2 & 5.125 & 5.132 & $\infty$ & 4.879 \\ 
3 & 7.924 & 7.934 & 11.62 & 17.51 \\ 
4 & 10.98 & 11.00 & $\infty$ & 533.9 \\ 
5 & 14.37 & 14.40 & 38.97 & 144.6 \\ 
6 & 18.13 & 18.17 & $\infty$ & 67.14 \\ 
7 & 22.31 & 22.35 & 95.15 & 123.4 \\ 
8 & 26.92 & 26.98 & $\infty$ & 1564 \\ 
9 & 32.00 & 32.07 & 197.3 & 965.4 \\ 
10 & 37.55 & 37.65 & $\infty$ & 303.9 \\ 
11 & 43.59 & 43.71 & 368.1 & 453.8
\end{tabular}
\end{center}
\caption{The natural frequencies $f_n^{(t)}$ predicted from theory~\cite{ari:2021,barbish:2022} and computed numerically $f_n^{(s)}$ where $n$ is the mode number. The variation of the theoretically predicted~\cite{ari:2021,barbish:2022} effective spring constant $k_n$ with mode number when measured at $x_0\!=\!1/2$ and $x_0\!=\!1/4$. The value of $k_n$ is a function of $x_0$ where $x_0$ is the location of both the application of the force and the measurement of beam displacement. Infinite values of $k_n$ occur when $x_0$ is at a node of the mode $n$.}  
\label{tab:fn-kn}
\end{table}

A location dependent effective mass, $m_n(x_0)$, and spring constant, $k_n(x_0)$, for each mode $n$ can be defined for a displacement measurement made at axial position $x_0$ along the beam.  This is accomplished by equating the kinetic and potential energy of the spatially extended beam with mode shape $\phi_n(x)$ with the kinetic and potential energy of the equivalent lumped mass attached to a spring~\cite{ari:2021,barbish:2022}. We emphasize, that the equivalent mass and spring constant include the role of tension through its influence on the mode shape.

We list the values of the effective spring constant, $k_n(x_0)$, in Table~\ref{tab:fn-kn}.  The stiffness of the modes $k_n$ increases with increasing mode number as expected. In addition, the effective spring constant depends strongly upon the location $x_0$ that is used in its definition. For example, if the location $x_0$ coincides with the location of a node of the mode of interest this results in an infinite effective spring constant at these locations. When the location is chosen to be the midpoint of the beam, $x_0\!=\!1/2$, only the modes with odd $n$ have finite values of the nodal spring constant since the midpoint is a node for all of the mode shapes with even values of $n$. The values of the effective masses can be found using $m_n \!=\! k_n/\omega_n^2$.

A convergence analysis of the static displacement of the beam due to a point force at the midpoint of the beam, $x_0\!=\!1/2$, yielded a similar conclusion with an error between theory and simulation of $\sim \! 0.2\%$. The theoretical prediction of the static displacement is computed using the sum of all contributing modes at the point $x_0$, where each mode $n$ has a spring constant $k_n(x_0)$. Overall, when the beam displacement is measured at $x_0\!=\!1/2$, the eleventh mode contributes $\sim \! 0.4\%$ of the overall displacement when considering the first eleven modes. For example, a 1 nm displacement of the beam at the midpoint would contain a contribution  of $\sim \! 4$ pm from the eleventh mode. 

We found that a numerical time step of $\Delta t \!=\! \tau_1/320$ was sufficient to compute the deterministic ring-down of the beam with sufficient resolution to describe the first eleven modes of the beam. The time scale $\tau_1$ is the period of oscillation of the fundamental mode of the beam in the absence of the fluid, $\tau_1 \!=\! f_1^{-1}$. This time step, $\Delta t \!=\! \tau_1/320 \!=\! 1.2$ ns, corresponds to a sampling frequency $f_s \!=\! 832$ MHz. The highest frequency we are interested in is the eleventh mode of the beam when immersed in air $f_{11} \!\approx\! 43~\text{MHz}$. Using our time step this results in $\sim \! 20$ time steps per oscillation of the eleventh mode which we found to yield sufficient time resolution.

The quality factors of the different modes of oscillation, $Q_n$, are mode dependent and depend significantly upon the surrounding fluid that is used. For a  ring-down simulation, a rough estimate of $Q_1$ is simply the number of oscillations that occur before the displacement becomes nearly zero in the beam's exponential approach to the stationary equilibrium state.  For a beam in a fluid, $Q_1$ (and more generally $Q_n$), is much smaller when the beam is immersed in water than when it is immersed in air. As a result, the total number of time steps required to resolve the beam's ring-down is significantly different depending upon the fluid. Therefore, the computational expense of the calculation increases significantly with increasing $Q_1$ due to the higher frequency dynamics and the longer time for the system to return to equilibrium.  Using a time step chosen to resolve eleven modes required a total number of time steps $N$ to compute the ring-down of $N\!=\!3,200$ in water and $N\!=\!12,800$ in air.  

\section{Results and Discussion}
\label{sec:results}

We now discuss the results of our computational study of the stochastic dynamics of the first eleven modes of the beam. We discuss the beam dynamics when it is immersed in water in Sec.~\ref{section:water} and when it is immersed in air in Sec.~\ref{section:air}. We consider two different computational domains where we vary the distance between the bottom surface of the beam and the floor below, $l_f$. In one case, $l_f \!=\! l_w$, and the presence of the floor is not expected to contribute significantly to the beam dynamics and these results are representative of a beam in an unbounded fluid. In the second case, $l_f \!<\! l_w$, and the floor is expected to affect the beam dynamics through viscous fluid interactions.

We compute the autocorrelation and the noise spectrum of the beam deflections at two axial locations along the beam, at the mid-point, $x_0\!=\!1/2$, and at the quarter-point, $x_0\!=\!1/4$. When the midpoint is used, only the modes with odd $n$ contribute which are all antinodes at this axial position. The modes with even $n$ all have nodes at the midpoint and therefore have zero displacement. Using the midpoint yields a very clean description of the dynamics of the modes with odd values of $n$. However, when the quarter-point is used, all of the eleven modes contribute to the dynamics. In this case, the computational approach captures the intricate and subtle features of the multimodal dynamics of all eleven modes.

\subsection{The Multimodal Dynamics of the Beam Immersed in Water}
\label{section:water}

We first examine the multimodal dynamics of the beam in water by  computing the autocorrelation and noise spectrum when the displacements are measured at the midpoint, $x_0 \! = \! 1/2$, and the floor is at a distance of $l_f\!=\!l_w$. The autocorrelation of equilibrium fluctuations in beam displacement are computed from the deterministic ring-down using Eq.~(\ref{eq:auto}). Prior to computing the ring-down, we initially deflect the beam by applying a point force of magnitude $F_0$ at $x_0$ and determine the stationary beam deflection. This deflected state of the beam is then used as the initial condition for a simulation without the externally applied force to yield the ring-down $W(x_0,t)$ for $t \!\ge\! 0$.

The autocorrelation is independent of the magnitude of $F_0$ used in the deterministic calculation as long as the beam displacements are small, where a Hookean spring description is valid, and the Reynolds number of the deterministic fluid dynamics remains small, such that the convective nonlinearity of the fluid equations is negligible~\cite{paul_stochastic_2006}. A good heuristic to follow when choosing the magnitude of $F_0$ to use in a numerical simulation, such that the beam remains Hookean and the fluid velocity based Reynolds number remains small, is to select $F_0$ such that $W(x,t) \!\ll\! h$ along the entire length of the beam. We have used $F_0$ such that $W(x_0,t \!=\!0) \! \sim \! 1$ nm which satisfies these requirements. 

\begin{figure}[h!]
\begin{center}
\includegraphics[width=3.1in]{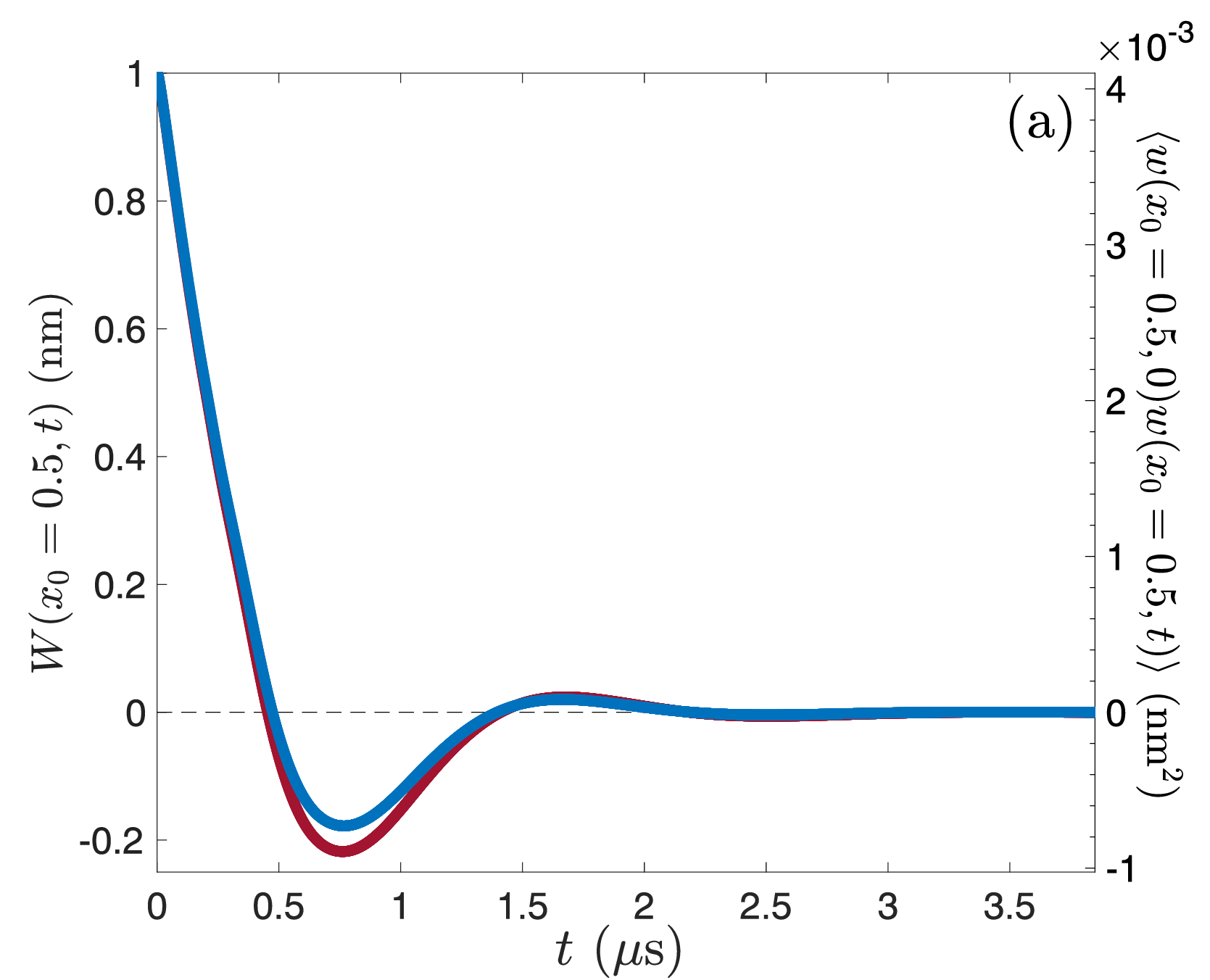}
\includegraphics[width=3.1in]{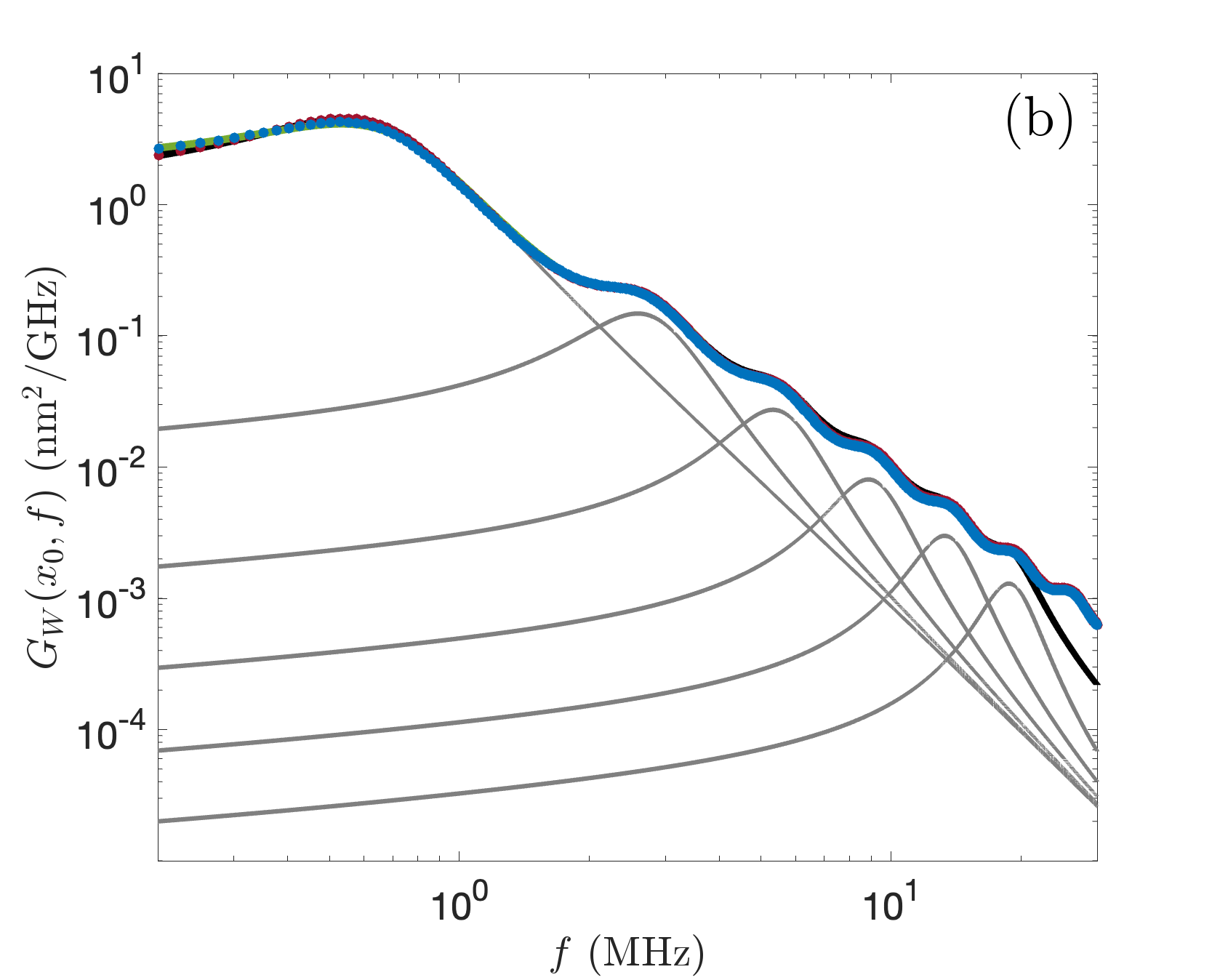}
\end{center}
\caption{The ring-down and autocorrelation~(a), and noise spectrum~(b), of the beam when immersed in water and measured at $x_0\!=\!1/2$ for two cases of floor separation: $l_f\!=\!l_w =5 \mu \text{m}$ (red), and $l_f \!=\! 2 \mu \text{m}$ (blue). (a)~Numerically computed ring-down (left axis), $W(x_0,t)$, and autocorrelation of equilibrium fluctuations $\langle w(x_0,0) w(x_0,t) \rangle$ (right axis) that is computed using Eq.~(\ref{eq:auto}). The horizontal dashed line is for reference. (b)~The noise spectrum $G_W(x_0,f)$ is computed using the autocorrelation from~(a) in Eq.~(\ref{eq:noise}). The black line is the theoretical prediction for a beam in an unbounded fluid~\cite{gress_multimode_2023} given by Eq.~(\ref{eq:gx}) and using only the first eleven modes. Gray lines are contributions from the individual modes with odd $n$ in the theoretical prediction for $n$ from one to eleven, the peaks appear in order from left to right. The sum of the gray curves yields the black curve. The green line is a semianalytical~\cite{tung_hydrodynamic_2008} prediction of the noise spectrum that accounts for the presence of the floor at a distance of $l_f \!=\! 2 \mu \text{m}$.}
\label{fig:water-x0p5}
\end{figure}

The ring-down, $W(x_0,t)$, and the autocorrelation, $\langle w(x_0,0) w(x_0,t) \rangle$, are shown in Fig.~\ref{fig:water-x0p5}(a) by the red curve using the left and right axis labels, respectively. As expected, the dynamics are strongly damped where the quality factor of the fundamental mode is $Q_1 \! \approx \! 1$. For $t \!\approx\!  10 \tau_1$ the deterministic beam deflection has reduced significantly to $W(x_0,t \! = \! 10\tau_0)/W(x_0, 0) \sim 10^{-4}$.  The influence of the higher modes are not clearly evident in the autocorrelation shown in Fig.~\ref{fig:water-x0p5}(a).

However, the multimodal contributions to the dynamics are clearly evident in the frequency representation of the dynamics as shown by the noise spectrum, see the red line in Fig.~\ref{fig:water-x0p5}(b). The noise spectrum, $G_W(x_0,f)$, is computed using the ring-down $W(x_0,t)$, shown in Fig.~\ref{fig:water-x0p5}(a), in Eq.~(\ref{eq:noise}). The noise spectrum contains a peak for each of the modes with odd $n$ from one to eleven when going from left to right.

A theoretical prediction for the noise spectrum is given by~\cite{gress_multimode_2023}
\begin{equation}
      G_W(x_0, \omega) = 4 k_B T \sum_{n=1}^{\infty} \frac{1}{k_n(x_0) \omega_n} \, \frac{\tilde{\omega}_n  T_0 \Gamma''(\omega)}{\left[1 - \tilde{\omega}_n^2 (1 + T_0 \Gamma'(\omega)) \right]^2 + \left[  \tilde{\omega}_n^2  T_0 \Gamma''(\omega)\right]^2}
\label{eq:gx} 
\end{equation}
where $k_n(x_0)$ is the mode and position dependent spring constant $k_n(x_0) \!=\! m \omega_n^2/\phi_n^2(x_0)$, $T_0 \!=\! \frac{\pi \rho_f b}{4\rho_s h}$ is the nondimensional mass loading parameter, $\tilde{\omega}_n \!=\! \omega/\omega_n$, and $\phi_n(x)$ is the mode shape of the beam for mode $n$. $\Gamma(\omega)$ is the complex-valued hydrodynamic function~\cite{sader_frequency_1998} that describes the frequency dependent fluid motion caused by the oscillating beam~\cite{rosenhead_laminar_1963,sader_frequency_1998,paul_stochastic_2006}.  The real part, $\Gamma'(\omega)$, describes the frequency dependent mass loading of the fluid. In general, the mass loading increases with decreasing frequency. The can be traced to the increasing value of the Stokes length with decreasing frequency through its $\omega^{-1/2}$ dependence. The imaginary part,  $\Gamma''(\omega)$, describes the frequency dependent damping of the fluid which monotonically decreases as the frequency is reduced~\cite{rosenhead_laminar_1963}. The mode shapes $\phi_n(x)$ are computed for an Euler-Bernoulli beam under tension using the approach~\cite{bokaian_natural_1990} described in detail using our notation in Refs.~\cite{barbish_dynamics_2022,ari:2021}.

Therefore, the appropriate form of $\Gamma(\omega)$ depends upon the geometry of the oscillating object, the details of the domain bounding the fluid, and basic assumptions of the important features of the flow field.  The flow field caused  by a long and slender beam in an unbounded fluid has been successfully modeled by neglecting axial flows and considering only the flow over the two-dimensional transverse cross-section of the beam.  For low Reynolds number flows, representing the rectangular cross section as a circular cylinder~\cite{rosenhead_laminar_1963} of diameter $b$ leads to analytical expressions that have been widely used to describe the dynamics of oscillating beams in fluid~\cite{sader_frequency_1998}.  The circular cylinder based hydrodynamic function can be corrected to account for the rectangular cross section of actual beams for a more accurate description using a thin blade approximation, as we show below, at the expense of more complicated expressions~\cite{sader_frequency_1998}. Brumley~\emph{et al.}~\cite{brumley:2010} have shown that for beams with finite rectangular cross-sections, the error in the blade approximation does not lead to significant errors and the blade description remains a very good approximation for flexural oscillations of beams over a wide range of conditions. When the axial flows are neglected, the resulting expressions are a function of frequency only and do not depend upon the mode number. An analytical expression for the hydrodynamic function of an oscillating blade of infinite length, finite width, and zero thickness that is immersed in an unbounded fluid has been developed by Van Eysden and Sader~\cite{van_eysden_frequency_2007} that accounts for the axial flows to yield a mode dependence. 

The black curve in Fig.~\ref{fig:water-x0p5}(b) is the theoretical prediction given by Eq.~(\ref{eq:gx}) using the mode independent $\Gamma(\omega)$ for a thin blade in an unbounded fluid. The hydrodynamic function for a thin blade can be expressed as~\cite{sader_frequency_1998}
\begin{equation}
\Gamma(\omega) = \Omega_c(\omega) \Gamma_c(\omega)
\label{eq:blade}
\end{equation}
where $\Gamma_c(\omega)$ is the hydrodynamic function for an oscillating cylinder of diameter equal to the beam width $b$ and $\Omega_c(\omega)$ is a complex-valued and frequency-dependent correction factor. The hydrodynamic function of an oscillating cylinder~\cite{rosenhead_laminar_1963,sader_frequency_1998} is
\begin{equation}
    \Gamma_c(\omega) = 1 + \frac{4 i \text{K}_1(-i \sqrt{i \text{Re}_\omega})}{\sqrt{i \text{Re}_\omega} \text{K}_0 (-i \sqrt{i \text{Re}_\omega})}
\end{equation}
where $i=\sqrt{-1}$ and $\text{Re}_\omega$ is the frequency based Reynolds number
\begin{equation}
    \text{Re}_\omega = \frac{\rho_f \omega b^2}{4 \mu_f}
\end{equation}
which can be thought of as a nondimensional frequency parameter. $\text{K}_0$ and $\text{K}_1$ are the zeroth and first order modified Bessel functions of the second kind, respectively. An explicit expression for the correction factor $\Omega_c(\omega)$ is provided in Ref.~\cite{sader_frequency_1998}.  The gray lines in Fig.~\ref{fig:water-x0p5}(b) are the contributions from each of the individual terms in the summation of Eq.~(\ref{eq:gx}) for odd $n$ between one and eleven.  The agreement between the theoretical prediction (black) and the numerical simulation for $l_f \!=\! l_w$ (red) is excellent for all of the modes shown.

Several interesting conclusions can be drawn from this comparison.  These results suggest that the mode independent hydrodynamic function, given by Eq.~(\ref{eq:blade}) and used in Eq.~(\ref{eq:gx}), accurately describes the fluid dynamics for the entire range of multimodal dynamics we explore for the flexural oscillations of a beam in fluid.  This also indicates that a more sophisticated hydrodynamic function with a modal dependence due to axial flows~\cite{van_eysden_frequency_2007} is not needed for the conditions we explore.  This is expected since the unsteady viscous boundary layers for the higher modes remain smaller than the modal wavelengths of the beam for the conditions we investigate.  In addition, these results also  indicate the accuracy of the assumption of independent modal contributions for the Brownian driven motion of the beam, as expressed by the summation in Eq.~(\ref{eq:gx}), which can be summed together to yield the final result.  These results show that our domain with $l_f \!=\! l_w$ is a good approximation of an unbounded fluid domain.

We next explore the influence of the proximity of the floor of the fluid domain, when placed closer to the beam at $l_f \!=\!2 \mu \text{m} \!<\! l_w$, on the resulting autocorrelation and noise spectrum.   The blue curve of Fig.~\ref{fig:water-x0p5}(a) shows the autocorrelation for $l_f \!=\!2 \mu \text{m}$ which results in additional damping.   The blue curve of the deterministic ring-down only reaches $-0.18$ nm at its first anti-node, while the red curve reaches $-0.22$ nm. The increase in damping is due to a squeeze-film damping mechanism which causes the fluid velocity gradient at the surface of the beam to increase due to the fluid having to move between the floor and the beam where viscous interactions are significant.

A semianalytical theoretical description of the hydrodynamic function $\Gamma(\omega)$ that accounts for the influence of a nearby rigid and no-slip bounding surface has been developed by Tung~\emph{et al.}~\cite{tung_hydrodynamic_2008}. The semianalytical expression captures the variation of the hydrodynamic function over a wide range of $l_f$ and $\text{Re}_\omega$. The hydrodynamic function is expressed as a fit to numerical results in terms of $\text{Re}_\text{L} \!=\! \log_{10}(\text{Re}_\omega)$ and $H_\text{L} \!=\! \log_{10}(H)$ where $H \!=\! 2l_f/b$ is the nondimensional distance to the floor.  An explicit expression is provided for $\Gamma_L \!=\! \Gamma_L(\text{Re}_L, H_L)$ in Table II of Ref.~\cite{tung_hydrodynamic_2008}. Lastly, the real and imaginary parts of the hydrodynamic function are determined as $\Gamma' \!=\!10^{\Gamma_L'}$ and $\Gamma'' \!=\! 10^{\Gamma_L''}$, respectively.

The green line in Fig.~\ref{fig:water-x0p5}(b) is the semianalytical prediction~\cite{tung_hydrodynamic_2008} when the floor is located at $l_f \!=\! 2 \mu \text{m}$. The semianalytical prediction for the noise spectrum is found using  Eq.~\eqref{eq:gx} with the the hydrodynamic function $\Gamma(\omega)$ developed by Tung \emph{et al.}~\cite{tung_hydrodynamic_2008}. The agreement between the semianalytical prediction and the numerical results is excellent and clearly captures the deviations from the unbounded fluid results at low frequencies. As expected, the floor does not influence the higher mode dynamics due to the higher frequencies of oscillation and the reduction in the unsteady viscous boundary layer thicknesses.

\begin{figure}[h!]
\begin{center}
\includegraphics[width=3.25in]{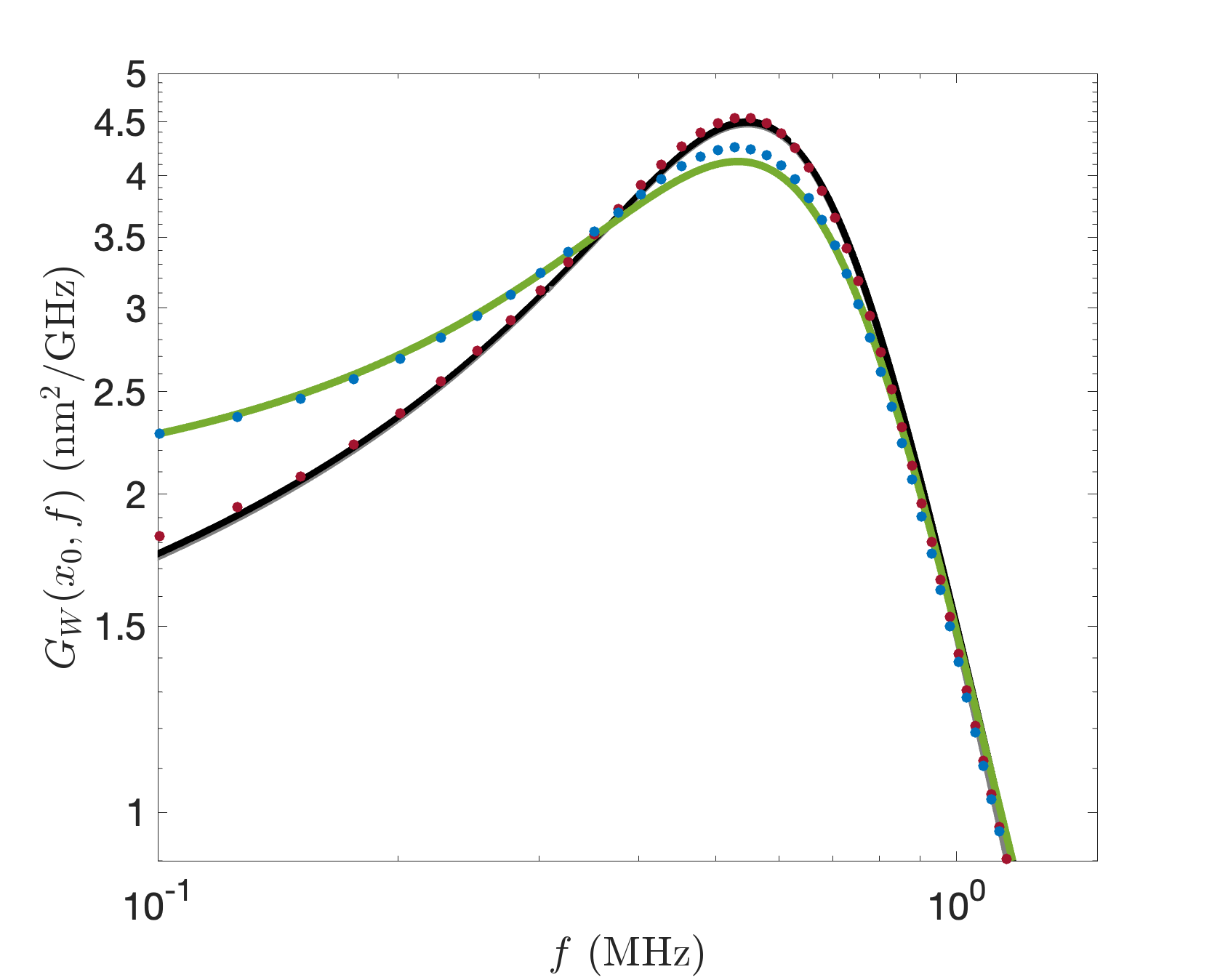}
\end{center}
\caption{A close-up view of $G_W(x_0\!=\!1/2,f)$ from Fig.~\ref{fig:water-x0p5} for lower frequencies illustrating the influence of the floor on the noise spectrum. The black line is the theoretical prediction for a beam in an unbounded fluid, the red symbols are numerical results for $l_f\!=\!l_w \!=\! 5 \mu \text{m}$, the green is curve is the semianalytical prediction~\cite{tung_hydrodynamic_2008} that accounts for the floor, and the blue symbols are numerical results for $l_f \!=\! 2 \mu\text{m} \!<\! l_w$.}
\label{fig:water-x0p5-closeup}
\end{figure}

In order to better visualize the effect of the floor of the fluid domain on the noise spectrum we show a close-up view emphasizing the lower frequencies in Fig.~\ref{fig:water-x0p5-closeup}. When the floor is further away at $l_f \!=\! l_w$ the agreement between the numerics (red symbols) and the theoretical prediction for an unbounded fluid is excellent.  However, when the distance to the floor is closer, $l_f \!<\! l_w$, the numerical results, shown by the blue symbols, deviate  from the unbounded case shown using the red symbols. The effect of the closer floor is only evident at frequencies near and below the peak frequency of the fundamental mode. These deviations from the unbounded results at lower frequencies are due to the increasing size of the unsteady viscous boundary layer with decreasing frequency as quantified by the Stokes length.

\begin{figure}[h!]
\begin{center}
\includegraphics[width=3.1in]{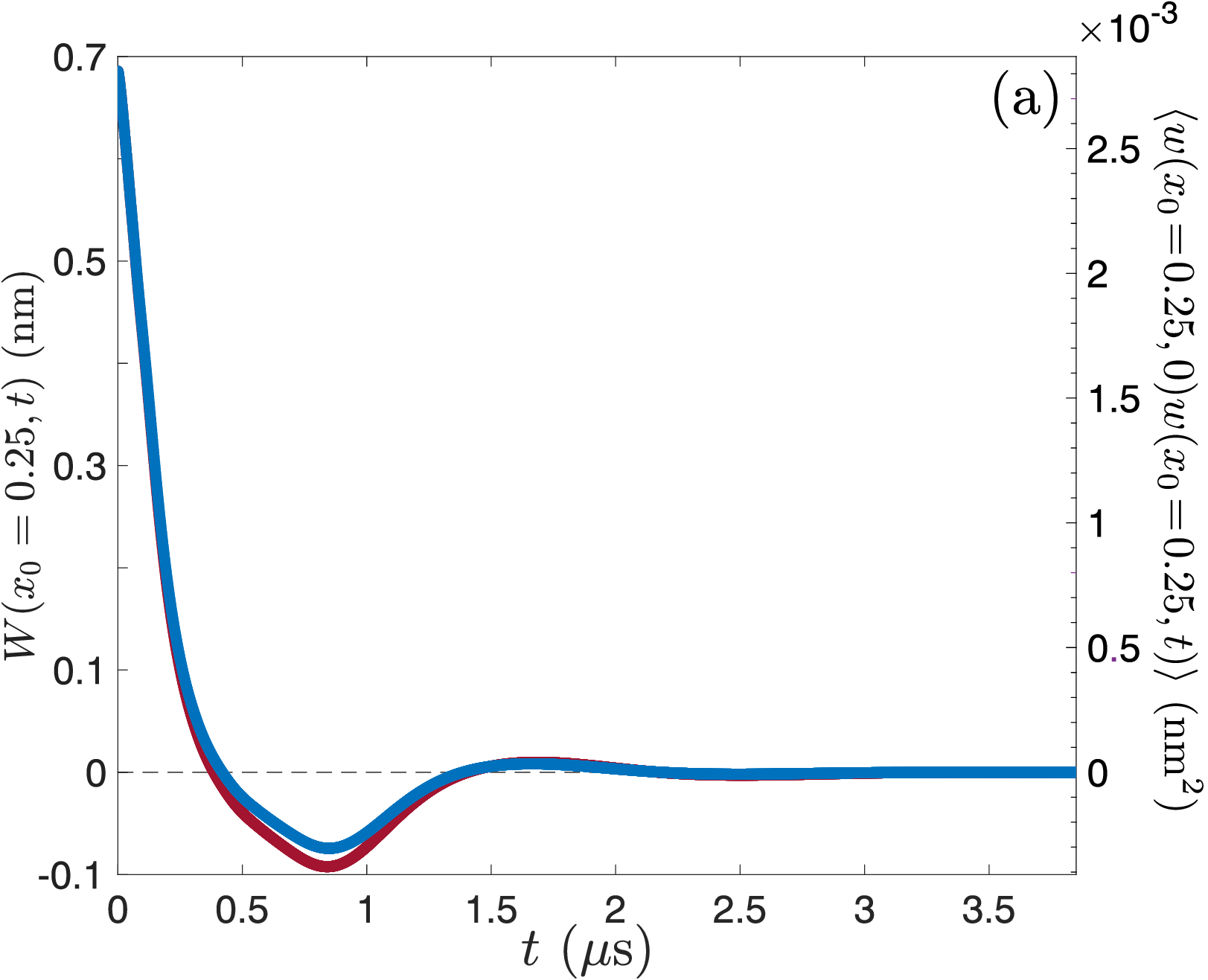}
\includegraphics[width=3.3in]{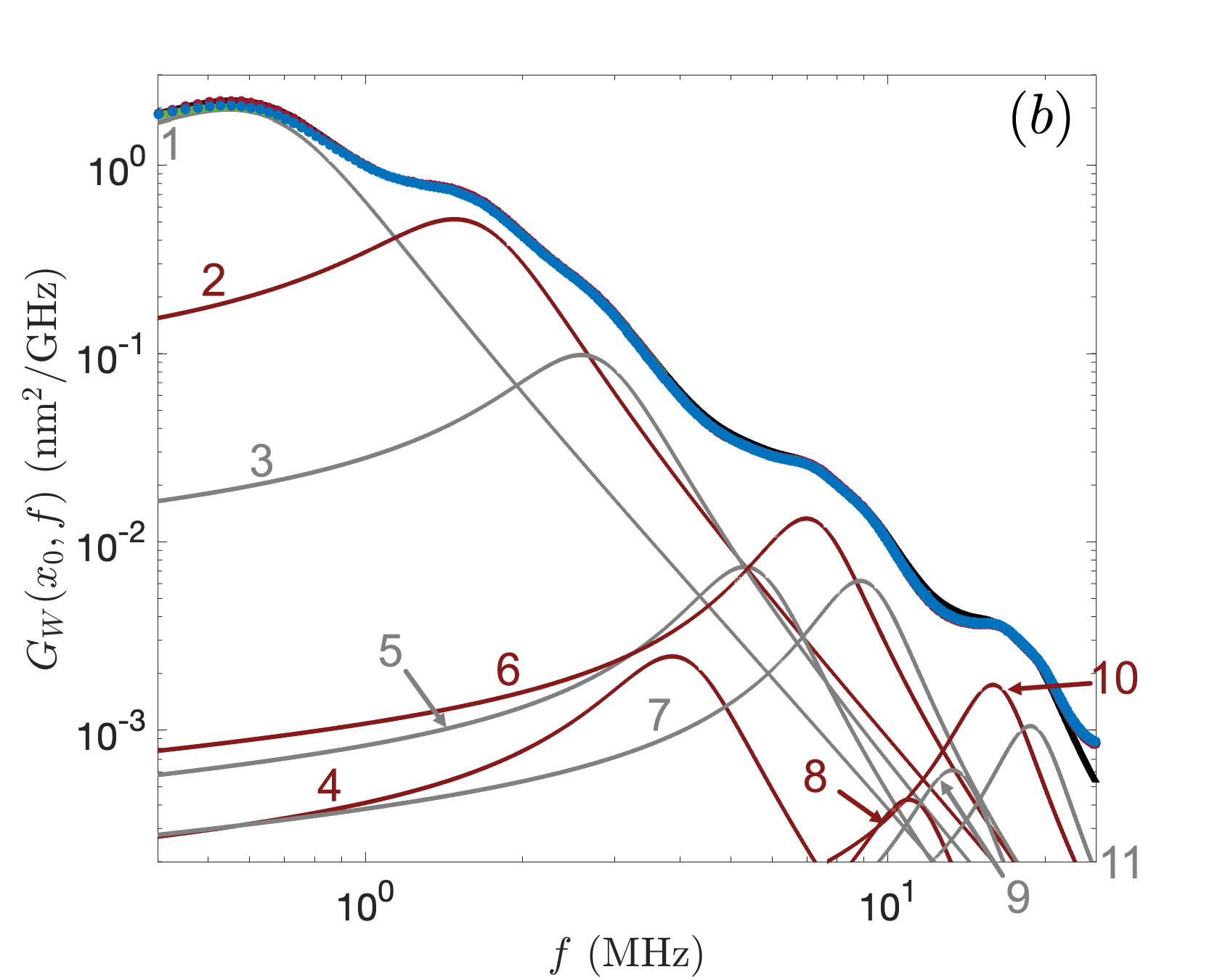}
\end{center}
\caption{The ring-down and autocorrelation~(a), and noise spectrum~(b), of the beam immersed in water when measured at $x_0\!=\!1/4$ where $l_f \!=\! l_w$ (red) and $l_f \!=\! 2 \mu \text{m}$ (blue).  (b) Theoretical prediction in an unbounded fluid (black) given by Eq.~(\ref{eq:gx}), individual modal contributions (labeled) from theory for odd $n$ (gray) and even $n$. The semi-analytical prediction~\cite{tung_hydrodynamic_2008} that accounts for the floor is shown in green.}
\label{fig:water-x0p25}
\end{figure}

The spectral properties of the stochastic beam dynamics are significantly different if the measurement of the deflection is made at $x_0 \!=\! 1/4$ as shown in Fig.~\ref{fig:water-x0p25}. In this case, modes of the beam with both even and odd $n$ contribute. This demonstrates how the computational approach can be used for any point of interest on the beam, and that it can capture the complex and overlapping contributions from multiple modes simultaneously. Following the same procedure, we compute the ring-down of the beam due to a force that was applied at $x_0 \!=\! 1/4$ in the distant past and is then removed at $t\!=\!0$. Using Eqs.~\eqref{eq:auto}-\eqref{eq:noise}, we compute the autocorrelation and noise spectrum.

A comparison of the numerical results with theoretical predictions for $x\!=\!1/4$ is shown in Fig.~\ref{fig:water-x0p25} using the same conventions as Fig.~\ref{fig:water-x0p5} where we show results for $l_f \!=\! 5 \mu \text{m}$ (red) and $l_f \!=\! 2 \mu \text{m}$ (blue). The ring-down and autocorrelation, shown in Fig.~\ref{fig:water-x0p25}(a), again yield a strongly damped return to equilibrium. However, in this case some deviation from the general shape of a simple harmonic oscillator's ring-down is evident as a result of the multimodal contributions. As expected, when the floor of the fluid domain is closer the increased fluid damping reduces the magnitude of the ring-down as it returns to equilibrium.

The noise spectrum spectrum is shown in Fig.~\ref{fig:water-x0p25}(b) where the gray curves are contributions from the modes with odd $n$, the red curves are contributions from modes with even $n$, and the sum of the modal contributions yields the theoretical prediction indicated by the black curve. In this case the modal contributions are quite complex where the peak amplitudes of the different modes are not monotonic with the mode number. The overall agreement between the theoretical prediction and the numerical results is excellent. It is clear that the location of the floor has very little influence upon the noise spectrum for these conditions, with only a minor shift for the first mode similar to the change in $G_W(x_0\!=\!1/2, f)$ shown in Fig.~\ref{fig:water-x0p5-closeup}. Included as the green curve is the semianalytical prediction that accounts for the floor which also aligns with the numerical and theoretical results that are shown.

\subsection{The Multimodal Dynamics of the Beam Immersed in Air}
\label{section:air}
We next explore the multimodal stochastic dynamics of the beam when immersed in air. In this case, the computations are much more expensive due to the high quality oscillations of the deterministic ring-down calculation. We chose to measure the displacement at $x_0\!=\!1/4$ which includes significant contributions from all eleven modes. We again explore two fluid domains,  in one domain $l_f\!=\!l_w\!=\!10 \mu$m and in the other domain $\l_f \!=\! 2 \mu \text{m}$.  For both computations, the simulation duration was for a total time of 40$\tau_0$, at which point the beam deflection at $x_0$ had reduced significantly to $W(x_0, t)/W(x_0, 0) \!\sim\! 5 \!\times\! 10^{-3}$.

\begin{figure}[h!]
\begin{center}
\includegraphics[width=3.0in]{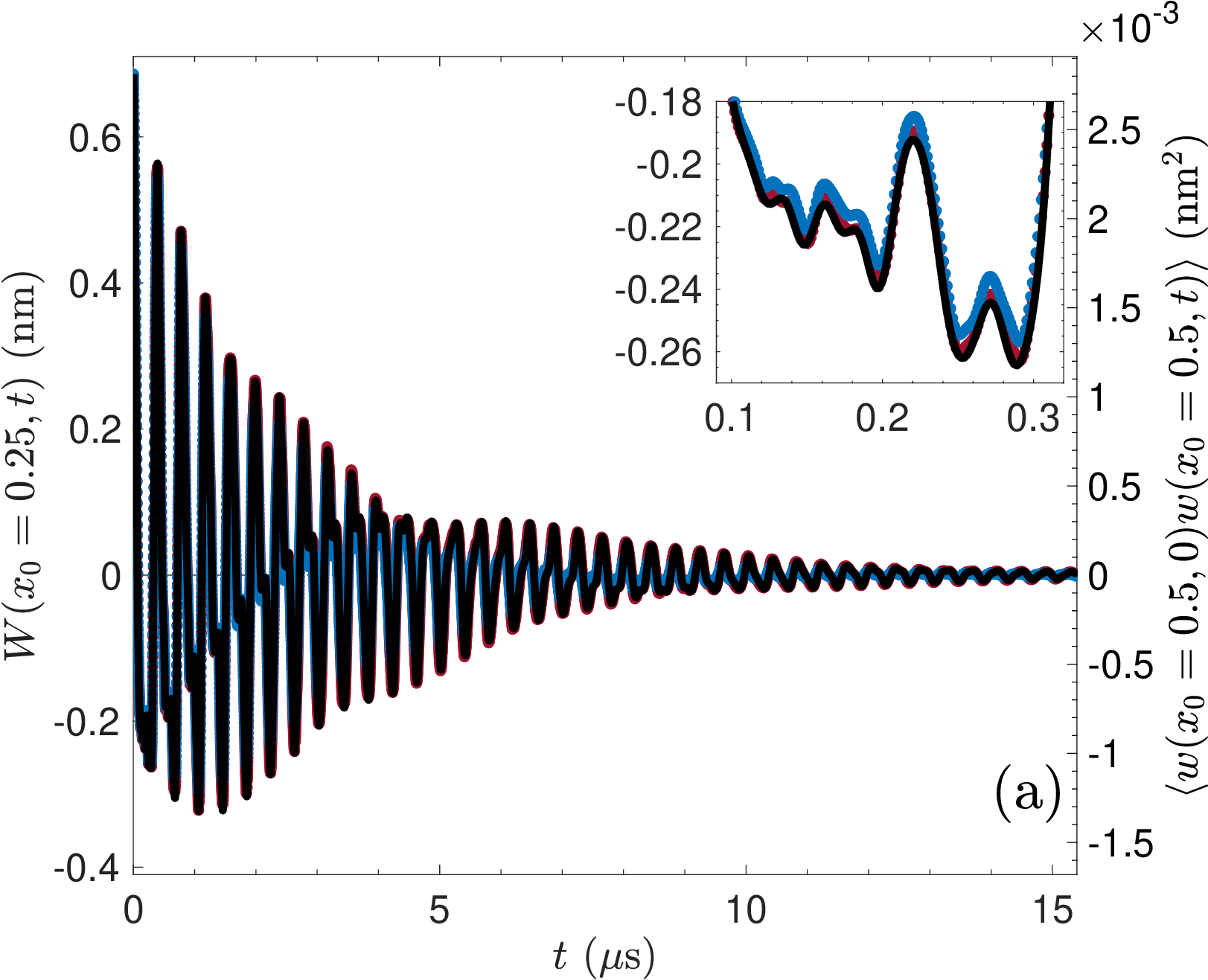} 
\includegraphics[width=3.4in]{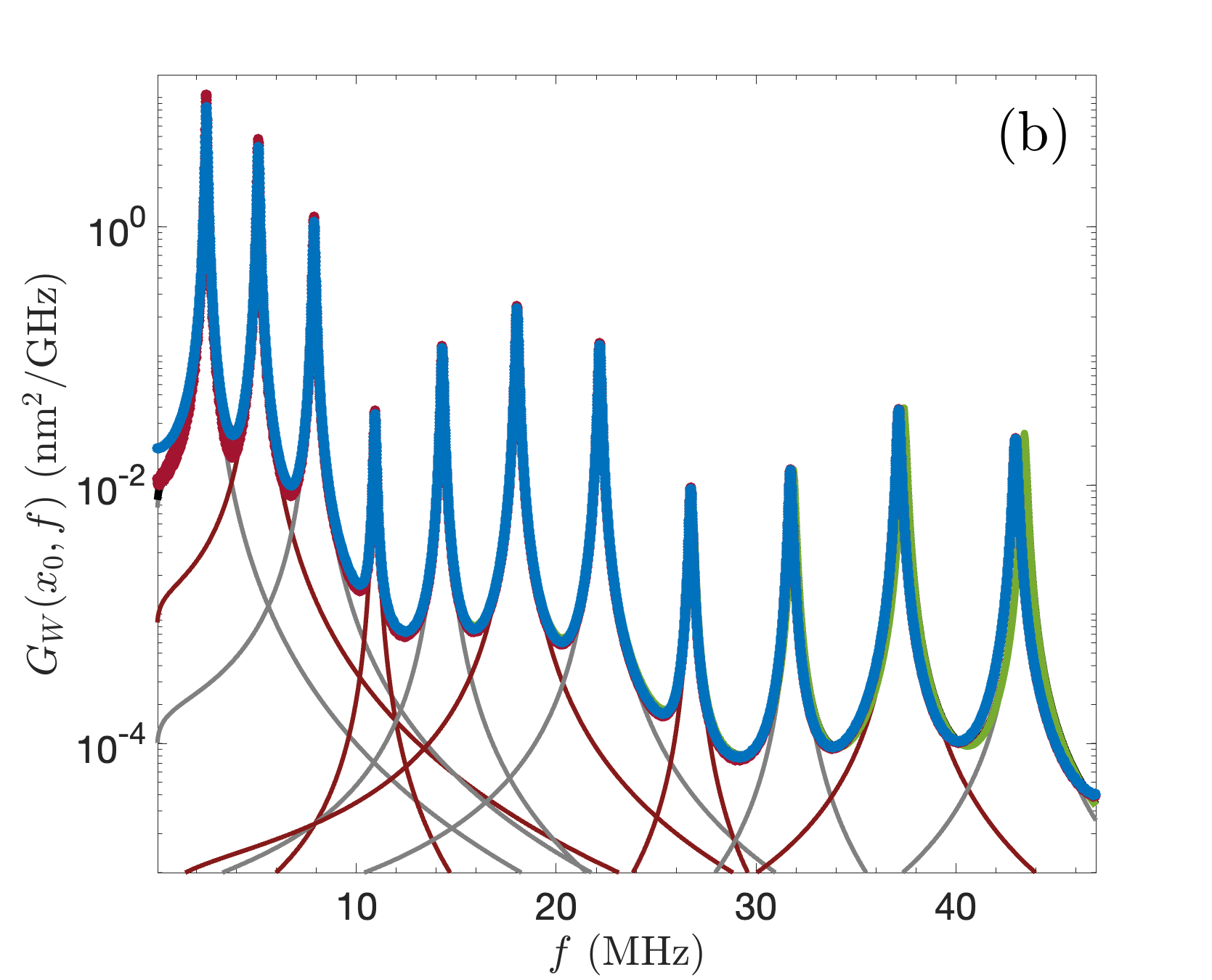}
\end{center}
\caption{The ring-down and autocorrelation~(a), and noise spectrum~(b), of the beam when immersed in air and the displacement the displacement is measured at $x_0 \!=\! 1/4$ using the same color conventions as Fig.~\ref{fig:water-x0p5}. Computational results for $W(x_0\!=\! 1/4, t)$ with $l_f \!=\! 10$ $\mu$m (red) and $l_f\!=\!2$ $\mu$m (blue). (a)~Inset shows a  closeup of the excellent agreement between theory and numerical simulation for these complex multimodal dynamics. (b)~The noise spectrum, eleven peaks are visible with increasing mode number from left to right.}
\label{fig:ring-down-beam-air-x0p25}
\end{figure}

The deterministic ring-down and autocorrelation are shown in Fig.~\ref{fig:ring-down-beam-air-x0p25}(a). The red and blue lines correspond to simulations with the floor at $l_f\!=\!10$ $\mu$m and $l_f \!=\!2$ $\mu$m, respectively. The time-domain dynamics of the beam are now quite complicated when compared to the highly damped results that were obtained when the beam was immersed in water.  In this case, the quality of the fundamental mode is $Q_1 \!=\! 26.5$.  The theoretical prediction for the ring-down, and autocorrelation, for the beam in an unbounded fluid are shown by the black curve. The theoretical prediction is found as the inverse cosine Fourier transform of the noise spectrum, given by Eq.~(\ref{eq:gx}), where we have included contributions from the first eleven modes. The agreement between the numerics and theory is excellent. The inset of Fig.~\ref{fig:ring-down-beam-air-x0p25}(a) shows a closeup view of the agreement and clearly illustrates the complex time-domain dynamics.

The noise spectrum is shown in Fig.~\ref{fig:ring-down-beam-air-x0p25}(b) where peaks from all eleven modes are evident. Modes one through eleven occur in sequence from left to right. Following our convention, the theoretical prediction is included as a black line and the semianalytical prediction is included as the green line. The agreement with these predictions is excellent and the lines nearly overlap as drawn. For modes ten and eleven there is some error evident in the location of the peaks. 

When the beam is in air, there are well defined peaks for each of the eleven modes of the noise spectrum shown in Fig.~\ref{fig:ring-down-beam-air-x0p25}(b). We quantify the differences in the noise spectra using the frequency of the peak, $f_{p, n}$, and the magnitude of $G_{W, p, n}$ evaluated at the peak frequency for each mode $n$ for the first eleven modes. We emphasize that $f_{p,n} \!<\! f_n$ when the beam is immersed in a viscous fluid where $f_n$ is the natural frequency.  The peak frequencies for each mode $n$ are listed in Table~\ref{tab:air-x0p25-fn}. The theoretical values, $f_{n,p}^{(t)}$, are computed by determining the frequency of the peaks of $G_W$ using Eq.~(\ref{eq:gx}).  The relative error of the frequencies are low where $E_r \!=\! f_{p,n}^{(s)}/f_{p,n}^{(t)} \!-\! 1$ with a maximum of $E_r \!\approx\! 1$\%.
\setlength{\tabcolsep}{8pt}
\begin{table}[h!]
\begin{center}
\begin{tabular}{ c  |r r r || r r r} 
 $n$ & $f_{p,n}^{(t)}$  & $f_{p,n}^{(s)}$  & $E_r$ (\%)  &$f_{p,n}^{(t)}$  & $f_{p,n}^{(s)}$   & $E_r$ (\%) \\ 
  &  & $l_f=10 \mu$m &  &  & $l_f=2 \mu$m  & \\    
\hline \hline 
1  & 2.486  & 2.489  &  0.14    &2.492   & 2.494     &0.08\\
2  & 5.083  & 5.088  &  0.11    &5.090  & 5.093     &0.08\\
3  & 7.870  & 7.874   &  0.05    &7.874   & 7.879      &0.07\\
4  & 10.920  & 10.926  &  0.06    &10.924  & 10.931     &0.07\\
5  & 14.297  & 14.295  &  -0.02   &14.298  & 14.300     &0.01\\
6  & 18.047  & 18.036  &  -0.07   &18.048  & 18.036     &-0.07\\
7  & 22.209  & 22.174  &  -0.16   &22.209  & 22.174     &-0.16\\
8  & 26.810  & 26.729  &  -0.31   &26.809  & 26.729     &-0.30\\
9  & 31.870  & 31.717  &  -0.48   &31.868  & 31.717     &-0.48\\
10 & 37.406  & 37.137  &  -0.72   &37.404  & 37.137     &-0.72\\
11 & 43.430  & 42.985  &  -1.03   &43.426  & 42.985     &-1.02
\end{tabular}
\end{center}
\caption{The peak frequency, $f_{p,n}$ (measured in MHz), for each mode $n$ in the noise spectrum $G_W(x_0\!=\!1/4,f)$ for a beam immersed in air and the relative error for each mode, $E_r \!=\! f_{p,n}^{(s)}/f_{p,n}^{(t)} \!-\! 1$.  Included are values from theory, $f_{p,n}^{(t)}$, and simulation, $f_{p,n}^{(s)}$, for $l_f \!=\! 10 \mu \text{m}$ and $l_f \!=\! 2 \mu \text{m}$.}
\label{tab:air-x0p25-fn}
\end{table}

The peak values of the the noise spectrum magnitude for each mode $n$, $G_{W,p,n}$, are listed in Table~\ref{tab:air-x0p25-gwn}. Additionally, the relative error for each mode, $E_r \!=\! G_{W, p, n}^{(s)}/G_{W, p, n}^{(t)} \!-\! 1$, is included and has magnitudes ranging from $\!<\!1$\% to 11\%. We note that the absolute error, $E_a \!=\! G_{W,p,n}/\max(G_{W})$, is 2.8\% for the first mode and that $E_a \!\lesssim\! 1\%$ for the remaining modes. We expect that the error in the first mode would be further reduced by continuing the simulations for times  larger than the $40\tau_0$ we used which resulted in $W(x_0,40 \tau_0)$ with a magnitude of $\sim 0.5$\% of its initial value. It is interesting to note that modes 4 and 8 have large relative errors. This is because modes 4 and 8 are near a node at $x_0 \!=\! 1/4$ and are significantly stiffer than their neighboring modes, and as a result have a smaller, and harder to resolve, contribution to the noise spectrum. 
\setlength{\tabcolsep}{8pt}
\begin{table}[h!]
\begin{center}
\begin{tabular}{ c  |l l r  l || l l r} 
 $n$ & $G_{W,p,n}^{(t)}$ & $G_{W,p,n}^{(s)}$ &  $E_r$ (\%)  & ~ &$G_{W,p,n}^{(t)}$ & $G_{W,p,n}^{(s)}$ & $E_r$ (\%)\\ 
  &  & $l_f=10 \mu$m &  &  & & $l_f=2 \mu$m  & \\    
 \hline \hline 
1  & 10.787         & 10.485         & -2.80  &&8.562          & 8.528         &-0.40\\
2  & 4.743          & 4.763          &  0.44  &&4.114          & 4.184         & 1.71\\
3  & 1.160          & 1.194          &  2.95  &&1.054          & 1.090         & 3.47\\
4  & $3.506 \times 10^{-2}$ & $3.782 \times 10^{-2}$ & 7.85 && $3.294 \times 10^{-2}$ & $3.541 \times 10^{-2}$ & 7.50\\
5  & 0.116          & 0.121          &  4.30  &&0.111          & 0.115         & 3.33\\
6  & 0.230          & 0.245          &  6.78  &&0.224          & 0.235         & 4.98\\
7  & 0.116          & 0.126          &  8.18  &&0.115          & 0.122         & 5.92\\
8  & $8.638 \times 10^{-3}$ & $9.584 \times 10^{-3}$ & 10.95  && $8.631 \times 10^{-3}$ & $9.351 \times 10^{-3}$ & 8.35\\
9  & $1.304 \times 10^{-2}$ & $1.330 \times 10^{-2}$ &  1.99  && $1.314 \times 10^{-2}$ & $1.301 \times 10^{-2}$ & -1.05\\
10 & $3.885 \times 10^{-2}$ & $3.879 \times 10^{-2}$ & -0.15  && $3.946 \times 10^{-2}$ & $3.815 \times 10^{-2}$ & -3.32\\
11 & $2.454 \times 10^{-2}$ & $2.304 \times 10^{-2}$ & -6.13  && $2.508 \times 10^{-2}$ & $2.278 \times 10^{-2}$ & -9.21
\end{tabular}
\end{center}
\caption{The peak value of the noise spectrum, $G_{W,p,n}(x_0 \!=\! 1/4,f)$,  for each mode $n$ in units of nm$^2$/GHz and the relative  error, $E_r = G_{W,p,n}^{(s)}/G_{W,p,n}^{(t)} \!-\! 1$. We include the peaks from theory $G_{W,p,n}^{(t)}$ and simulation $G_{W,p,n}^{(s)}$ 
for $l_f \!=\! 10 \mu \text{m}$ and $l_f \!=\! 2 \mu \text{m}$.}
\label{tab:air-x0p25-gwn}
\end{table}

The influence of the presence of the floor, when the beam is immersed in air, again results in increased dissipation due to squeeze-film damping mechanisms. However, the overall impact is smaller than what was found using water as the surrounding fluid. When the floor is nearby, and the fluid is air, the amplitude of the noise spectrum decreases, as shown by the 18.7\% decrease in the first mode peak amplitude. As the frequency increases, the differences between the floor at $l_f\!=\!2$ $\mu$m and $l_f\!=\!10$ $\mu$m decrease as well, with only small differences of a few percent for the highest modes studied here as a result of the decreasing thickness of the unsteady viscous boundary layer with increasing frequencies of oscillation.

\section{Conclusions}
\label{sec:conclusions}
Overall, we have demonstrated how a single deterministic simulation can resolve the multimodal, stochastic, and flexural dynamics of a micro or nanoscale elastic object  immersed in viscous fluid. We have shown this by computing the autocorrelations and noise spectra for the first eleven modes of a  doubly-clamped beam under tension for experimental conditions where we have found excellent agreement with theoretical predictions.  We have also demonstrated the flexibility of the  approach by computing the autocorrelations and noise spectra of the stochastic beam displacements at two different axial positions along the beam. Although we used $x_0\!=\!1/2$ and $x_0\!=\!1/4$, it is straightforward to apply this approach to any point on the elastic object. The computational approach captures the complete multimodal stochastic dynamics of the beam immersed in fluid which includes the precise geometry of experiment, the three-dimensional fluid dynamics, and the full fluid coupling with surrounding walls or other elastic objects.

In our study of an experimentally relevant nanoscale beam, we found that a mode independent hydrodynamic function accurately described the multimodal flexural dynamics of the first eleven modes of the beam in air or water. Our results validate the use of analytical predictions for the flexural dynamics of a beam, up to a high mode number, that is immersed in a viscous fluid. More broadly, our results provide support that the computational approach can be extended further to explore more complex three-dimensional problems where analytical results are not available.  

Our computational study used readily available finite element methods that can numerically integrate the equations of motion for complex spatial domains containing fluids and solids with complicated and intricate fluid-solid interactions that often exist in the laboratory or are at the center of new technologies.  It is important to emphasize that the fluctuation-dissipation approach we describe is very general. In fact, the only assumptions are classical mechanics and linear response~\cite{paul_stochastic_2006}. We have explored a beam undergoing flexural oscillations in a viscous fluid and all of the analytical results we present are for this situation.  However, we emphasize that general numerical approach can be applied much more broadly.

In particular, the computational approach we describe and use can be readily extended to other forms of motion with different dissipative phenomena. For example, beam dynamics due to compression modes, torsional modes, in-plane bending, \emph{etc}. In addition, other forms of dissipation beyond that of a viscous fluid can be used as well. For example, a fluctuation-dissipation based approach is at the heart of the LIGO measurements of gravitational waves which measures, with great precision, the fluctuations of hanging test masses~\cite{saulson:1990}. 

The numerical approach can also be readily extended to the study of the multimodal dynamics of complex structures. Interesting examples include a nanoscale helical spring with an attached bead~\cite{ushiba_size_2015}, elastic bowtie structures and microscale combs~\cite{jayne_dynamic_2018}, and more recently 3D nanoprinted structures~\cite{stassi_reaching_2021}. This computational  approach could be directly applied to describe the stochastic dynamics of complex geometries, such as these, to provide insights into the fluid-solid interactions and multimodal stochastic dynamics of the next generation of micro and nanoscale structures.  

\section*{Acknowledgements}
The authors gratefully acknowledge support from the National Science Foundation (NSF) (Grant No. CMMI-2001559). We thank Hagen Gress and Kamil Ekinci for many fruitful and insightful discussions. Portions of the computations were conducted using the resources of the Advanced Research Computing (ARC) at Virginia Tech.

\section*{Data Availability}

The data that support the findings of this study are available from the corresponding author upon reasonable request.

\end{document}